\documentclass[10pt,twocolumn,preprintnumbers,amsmath,amssymb,nofootinbib
,superscriptaddress]{revtex4}
\usepackage{graphicx,longtable,mathrsfs,color,array}
\usepackage[hidelinks]{hyperref}
\usepackage[usenames,dvipsnames]{xcolor} 
\usepackage{amssymb,amsmath,mathtools,mathrsfs,slashed} 
\usepackage{epsfig,subfigure,placeins,float} 
\usepackage{booktabs,longtable,ctable,multirow} 
\usepackage{exscale,relsize} 
\usepackage[normalem]{ulem} 
\usepackage{enumerate}
\usepackage{times, mathptmx} 
\usepackage{color}
\usepackage{hyperref}
\usepackage{graphicx}
\usepackage{color}
\usepackage{graphicx,graphics}

\newcommand{\be}{\begin{equation}}  
\newcommand{\ee}{\end{equation}}  
\newcommand{\bear}{\begin{eqnarray}}  
\newcommand{\eear}{\end{eqnarray}}  
\newcommand{\ba}{\begin{array}}  
\newcommand{\ea}{\end{array}}

\newskip\humongous \humongous=0pt plus 1000pt minus 1000pt  
\def\caja{\mathsurround=0pt}  
\def\eqalign#1{\,\vcenter{\openup1\jot \caja  
	\ialign{\strut
\documentclass[10pt,twocolumn,preprintnumbers,amsmath,amssymb,nofootinbib,superscriptaddress]{revtex4} \hfil$\displaystyle{##}$&$  
	\displaystyle{{}##}$\hfil\crcr#1\crcr}}\,}  
\newif\ifdtup

\def\oldreffmt#1{\rlap{[#1]} \hbox to 2\parindent{}}

\def\figfmt#1{\rlap{Figure {#1}} \hbox to 1in{}}  
  
\def\ie{\hbox{\it i.e.}{}}	  
\def\eg{\hbox{\it e.g.}{}}	  
\def\etal{\hbox{\it et al.}}  
  
\def\beq{\begin{equation}}  
\def\eeq{\end{equation}}  
\def\bea{\begin{eqnarray}}  
\def\eea{\end{eqnarray}}

\def\bq{\begin{quote}}  
\def\eq{\end{quote}}

\def \etal {{\it et al.}\ }  
\relax  

\newdimen\tdim  
\tdim=\unitlength  
\def\bar{\overline}

\begin{document}

\preprint{FERMILAB-PUB-15-438-T}

\title{Reply to 
``Comment on `Axion induced oscillating electric dipole moments' ''
}

\author{Christopher T. Hill}
\email{hill@fnal.gov}
\affiliation{Fermi National Accelerator Laboratory\\
P.O. Box 500, Batavia, Illinois 60510, USA}

\date{\today}

\begin{abstract}
We respond to a paper of Flambaum, \etal [Phys.\ Rev.\ D {\bf 95}, no. 5, 058701 (2017)], 
claiming there is no
effective induced oscillating electric dipole moment, \eg, for the electron, arising  
from interaction with an oscillating cosmic axion background via the anomaly.  
The relevant Feynman amplitude, Fig.(1), as computed by 
Flambaum \etal,  becomes a total divergence, and vanishes. 
Contrary to this result, we obtained
a nonvanishing amplitude,
that  yields physical electric dipole radiation 
for an electron (or any magnetic dipole moment) immersed in a cosmic axion field.
We argue that the Flambaum \etal counter-claim is incorrect,
and is based upon
a misunderstanding of a physics choice vs. gauge choice, and an assumption
that electric dipoles be defined only by coupling to 
{\em static} (constant in time) electric fields.

\vspace{0.2in}

\noindent
DOI: 10.1103/PhysRevD.95.058702 
\end{abstract}

\maketitle

\vspace{0.1in}


In recent papers \cite{Hill1,Hill2, Hill3} we have computed the effect of a coherent
oscillating axion dark matter field, via the electromagnetic anomaly, upon
the magnetic moment of an electron, or arbitrary magnetic multi-pole source.  
Figure (1) has been computed in several
ways and the results are consistent, nontrivial, and have potentially 
interesting physical and observational implications.

This can be viewed as a scattering amplitude for the coherent cosmic axion
field on a heavy, static, magnetic dipole moment, with conversion to an outgoing photon
or classical radiation field.  We find, however, that
this  leads to the consistent
interpretation that the electron behaves as though it has
acquired an ``effective oscillating
electric dipole moment'' (OEDM) in the background oscillating cosmic axion field,
which then acts as a source for electric dipole radiation.

In ref.\cite{Flambaum}, however,
it is claimed that the results of the analysis \cite{Hill1,Hill2} are wrong.
The authors actually claim  that the Feynman
diagram of Fig.(1) ``when properly computed''  vanishes.

We emphatically disagree with the conclusions of Flambaum, \etal
We show that they have made assumptions that lead them to  compute
a vanishing total divergence. Indeed, we previously computed the 
full effective action for a stationary
electron in an arbitrary gauge, \cite{Hill1,Hill2}.  One can
readily see that it contains the Flambaum \etal result in their special limit,
where  indeed it reduces to a vanishing total divergence.  
However, the full amplitude is nonvanishing and physical, and the Flambaum \etal limit
is irrelevant and misses the physics.

\vspace{0.1in}

Let us first review the situation. In the simplest case,
we consider the comoving cosmic axion field
$a(t)/f_a =\theta(t) = \theta_0\cos(m_a t)$, in the
limit of a stationary, non-recoiling electron (this is the relevant limit
since the axion mass  $m_a << m_e$). From Fig.(1)
we obtain   the following
effective interaction, written in terms of nonrelativistic two-component
spinors \cite{Hill1}:
\beq
\label{one}
\int d^4x\; g_a\;\mu_{Bohr}{\theta}(t) \psi^\dagger\vec{\sigma}\psi \cdot \vec{E} 
\eeq
This result is a contact term and is computed  in radiation gauge, where the electric field
is $\vec{E}=-\partial_t \vec{A}$ for vector potential $\vec{A}$
and $\vec{\nabla}\cdot\vec{A}=0$. In momentum space it takes the form
$ g_a m_a\;\mu_{Bohr}\theta_0 \psi^\dagger\vec{\sigma}\psi \cdot \vec{\epsilon} $
where $\vec{\epsilon} $ is the photon polarization. Clearly the amplitude
vanishes in the limit $m_a\rightarrow 0$.  The $m_a$ factor is absorbed into 
$\vec{E}=-\partial_t \vec{A}$ in writing eq.(\ref{one}).

\begin{figure}[tbp]
\vskip0.1in
\begin{center}
\includegraphics[width=4cm, height=4cm]{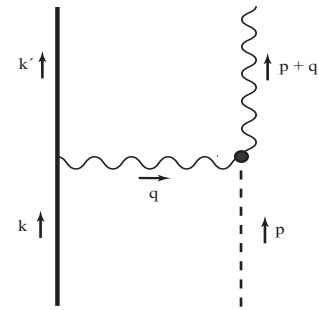} 
\end{center}
\caption{The dotted vertex is the axion-anomaly, $\theta F\widetilde{F}$,
and the solid vertical line is the electron. The 
electron--photon vertex is the magnetic moment of the electron.  The incoming
axion with 4-momentum $(m_a,\vec{0})$ absorbs a spacelike photon of 4-momentum 
$(0, \vec{k})$ with $|\vec{k}|=m_a$ to produce an outgoing photon
of momentum $\sim (m_a, \vec{k})$.  The electron barely recoils, 
since $m_e >> m_a$..}
\label{figure_back}
\end{figure}

Given the form of this result, we interpret this
as an effective, induced OEDM for the electron.  We claim this
result is general, and the interaction produces 
electric dipole radiation from
any static magnetic moment immersed in, and 
absorbing energy from, the oscillating cosmic axion field.
Indeed, since the result follows from a tree-diagram, it
can be demonstrated classically by a straightforward manipulation of
Maxwell's equations,  \cite{Hill3}. The 
radiation is formally that of an oscillating (Hertzian) 
electric dipole, with outgoing electric field polarization
aligned in the direction of the magnetic moment, and thus apparently violating CP.  
The emitted power by a free electron, in a spin-up to spin-up
transition,  is (for a derivation see section IV.B of \cite{Hill2}):
\bea
\label{two}
P & =& \frac{1}{12\pi} (g_{a}\theta_0\; \mu_{Bohr})^2 m^4_a 
\eea
This result is equivalent to that obtained from the classical Maxwell equations
for a fixed classical magnetic moment $\vec{m} = 2\mu_{Bohr}(\vec{s}/2)$ with a
spin unit-vector $\vec{s}$
\cite{Hill2,Hill3}.

More generally, we have computed  Fig.(1) in an arbitrary gauge for the
background electric field, \cite{Hill1, Hill2}.
We obtained in the static
 $ \vec{P}(x)= \mu_{Bohr}\psi^\dagger\vec{\sigma}\psi $ limit:
\beq
\label{three}
S=g\int d^{4}x\; \theta (t)\left( \vec{P}\cdot 
\vec{E}+\vec{\nabla }\cdot \vec{P}\left( 
\frac{1}{\vec{\nabla }^{2}}\right) \vec{\nabla }\cdot 
\vec{E}\right) 
\eeq
This result
differs from the radiation
gauge result eq.(\ref{one}) by the appearance of the nonlocal term.
Such nonlocal terms occur in electrodynamics when certain gauge choices are
specified, as in the case of the ``transverse current,'' (see below
and \cite{Jackson}).
Here, $\frac{1}{\vec{\nabla }^{2}}$ 
is a static Green's function,
\ie,  
\beq
\label{four}
 A(x)
\frac{1}{ \vec{\nabla }^{2} } B(x) 
=
 \int d^{4}y\; A(x)\frac{\delta(x_0 -y_0) }{ 4\pi | \vec{x}-\vec{y}| }B(y)
\eeq
In an arbitrary gauge,
$\vec{E}=\vec{\nabla }\varphi
-\partial _{t}\vec{A}$, after integrations by parts, 
the action of eq.(\ref{three})
takes the form:
\bea
\label{five}
&& S  =g\int d^{4}x\; \theta (t)\vec{\nabla }\cdot (
\vec{P}\varphi) 
\nonumber \\
& &\!\!\!\!\!\!\!\!  +g\int d^{4}x\; \partial _{t}\theta
(t)\left( \vec{P}\cdot \vec{A}+\vec{
\nabla }\cdot \vec{P}\left( \frac{1}{\vec{\nabla }^{2}}
\right) \vec{\nabla }\cdot \vec{A}\right) 
\eea
This result is indeed gauge invariant as can be checked explicitly,
as it is just a rewrite of the manifestly gauge invariant eq.(\ref{three}).
If there are no surface terms we can drop the 
first term on the {\em rhs} which is a total divergence,
and  with $\vec{\nabla }\cdot \vec{A}=0$ 
(radiation gauge; this follows from $\vec{\nabla }\cdot \vec{E}=0$
upon integrating by parts in time) the result reduces back to eq.(\ref{one}).
It should be noted that the first term on the {\em rhs} of 
eq.(\ref{three}) or eq.(\ref{five}) actually represents a force exerted
upon the OEDM by an applied oscillating $\vec{E}$,
hence there is potentially more physics here than dipole radiation.

\vspace{0.1in}

We can now see several
flaws with the Flambaum \etal analysis. They have ``properly computed'' 
this result in the particular case $\vec{A}=0$ and
$A_0=\varphi\neq 0$.  In this case we see that only the first term
will be formally nonzero in eq.(\ref{five}), but that term
is just a spatial total divergence, and
hence it contributes nothing to the physics. A total divergence 
is zero in momentum space and the Feynman
diagram of Fig.(1) then yields zero.  

Moreover, Flambaum \etal claim that this 
is a ``gauge choice.'' But this is, in fact, {\em a physics choice} since
one cannot generally make $\vec{A}$ vanish by a gauge transformation.
Furthermore, a time dependent $A_0=\varphi $  {\em necessarily requires a nonzero $\vec{A}$
by equations of motion} as we show in the discussion below
eq.(\ref{eight}).
Therefore,  Flambaum \etal, by
using only a Coulomb potential to probe a
dynamical time dependent radiating source, are forcing the external
field to be static and thus obtain
a false null result by Fourier mismatch, as well as total divergence.
Finally, their result is consistent with our result in taking
the pure Coulomb or static limit,  but it is
our result which they are attacking!

\vspace{0.1in}
 
The many conceptual errors and discrepancies of Flambaum \etal with our results seem to stem 
from a faulty definition  which they claim
to be valid for any EDM.   They state:

\vspace{0.1in}

``(1) The EDM of an elementary particle is defined by the linear energy shift that it produces
through its interaction with an applied {\em static} electric field: $\delta\epsilon = −\vec{d}\cdot\vec{E}$. 
As we show explicitly,
the interaction of an electron with an applied {\em static} electric field, in the presence of the axion
electromagnetic anomaly, in the lowest order does not produce an energy shift in the limit
$v/c \rightarrow  0$. This implies that no electron EDM is generated by this mechanism in the same
limit.''

\vspace{0.1in}

While this definition may be applicable to a static EDM, as in an
introductory course in electromagnetism,  
{\em it is inapplicable to an intrinsically  time dependent one.}
With an OEDM we are dealing
with a dynamical situation and must resort to a
more general definition, phrased in the context 
of an action.

We  should define the  EDM or OEDM of  any
object  as a covariant action of the form:
\beq
\label{six}
S = g\int d^4 x\;  S_{\mu\nu}(x) F^{\mu\nu}(x)
\eeq
where  $S_{\mu\nu}$ is an antisymmetric odd parity dipole density
 (\eg, $S_{\mu\nu} \sim \bar{\psi}\sigma_{\mu\nu} \gamma^5\psi$
for a relativisitic particle). 

For concreteness, let us consider the case
of the axion induced neutron OEDM. The neutron OEDM
is believed to arise in QCD from instantons.  It is being sought in a
proposed experiment (see ref.\cite{budkher} and references therein).
In the common rest frame of the neutron and axion,
the OEDM action of eq.(\ref{six}) reduces to:
\beq
\label{seven}
S=g\int d^{4}x\; \theta (t) \vec{P}\cdot 
\vec{E}(t) 
\eeq
where  $ \vec{P}(x)=(e/m_N) \psi^\dagger \vec{\sigma}\psi(x)$ 
is the dipole spin density,
written in terms of two-component spinors. 
$ \vec{P}(x)$ is localized in space
and static (time independent),
and the oscillating aspect of the EDM comes from the axion $\theta(t)$.

Note that a non-recoiling  neutron is the kinematically favored limit, 
\eg, as in Fig.(1). The neutron (or electron) is very heavy
compared to the axion, and like a truck being hit by a ping-pong ball
can only
acquire an insignificant kinetic energy. Therefore,
the radiated photon must carry off the full energy of the incident axion,
with a 4-momentum of $(m_a, \vec{k})$,  and $|\vec{k}|=m_a$
(and the exchange photon 4-momentum
is spacelike, $(0, \vec{k})$). 

Clearly, for a constant background 
electric field the actions of eqs.(\ref{one},\ref{seven}) average to zero. 
The radiated photon is necessarily 
time dependent with frequency $m_a$, as
will be the case for any OEDM.
In the case of a  radiation gauge photon, we have $A_0=0$ and 
a non-zero $\vec{A}$ with $\vec{\nabla}\cdot \vec{A}=0$. In this case our action
for the neutron OEDM is indistinguishable from the OEDM of the electron
of eq.(\ref{one}). Both 
require a time dependent $\vec{E}$, 
and are $\propto \partial_t\theta(t)$ upon integration 
by parts in time.

\vspace{0.1in}

Our result of eq.(\ref{one}), induced by the axion-QED anomaly,
has also been attacked by several other individuals 
for violating the Adler
decoupling of the axion.  The decoupling limit corresponds to $m_a\rightarrow 0$
and it superficially appears that eq.(\ref{one}) does not vanish in this limit as decoupling would
dictate (of course, it came from the momentum-space
result that was obviously $\propto m_a$, and this appears explictly
in ref\cite{Hill1}).  
However, in refs.\cite{Hill2,Hill3} the issue
of the axion decoupling is studied in detail, and it is
found to be somewhat subtle in general.  

In fact, eq.(\ref{one}) displays the same
behavior as the anomaly itself. The anomaly, in a constant
$\vec{B}$ field, can be written either in a manifestly
gauge invariant 
form $\propto \theta(t) \vec{E}\cdot \vec{B}$ or in a manifestly decoupling form
$\propto \partial_t (\theta(t)) \vec{A}\cdot \vec{B}$ where 
$\vec{E}=-\partial_t \vec{A}$
in a radiation gauge.  It is not possible to display simultaneously
the manifest decoupling, and gauge invariance. Likewise, in the static electron
limit eq.(\ref{one}) can be written as:
\beq
\label{eight}
\int d^4x\; g\;\mu_{Bohr}\partial_t{\theta}(t) \psi^\dagger\vec{\sigma}\psi \cdot\vec{A}
\eeq
where $\vec{A}$ is the vector potential. Here we see manifest decoupling, but
an expression written in terms of a vector potential. More generally
the result
in an arbitrary gauge with recoil can be derived and displays the same
behavior.

The decoupling is actually subtle and beautiful. One can see this explicitly in 
the eqs.(56,57) of ref.\cite{Hill2} for the near-zone radiation field 
(and in eqs.(44) for the RF cavity)
and in the classical analysis of \cite{Hill3}.  The decoupling is actually
occuring in the spatial structure of the nearzone radiation field
(or RF cavity modes). These vanish  as $m_a^2$
due to a ``magic cancellation:'' the static magnetic dipole field, 
which multiplies $\theta(t)$, does not radiate and
cancels, in the $m_a\rightarrow 0$ limit, against the outgoing radiation field 
which is retarded and proportional to $\theta(t-r/c)$, leaving
terms of order $m_a^2$. This implies that here there is no ``Witten effect,''
whereby a constant induced electric dipole would remain 
in the $\theta\rightarrow$ constant
limit: the
 would-be Witten term cancels against the retarded outgoing radiation field
in the near-zone.
In the end the radiated power is $\propto m_a^4$,
and axion decoupling is certainly working as it should.
Such radiation is physically
interesting, and may be detectable in experiment \cite{Hill2}. 

\vspace{0.1in}

Let us consider the problem of allowing $A_0$ to be time dependent while 
trying to maintain $\vec{A}=0$. 
$A_0$ is a non-propagating
field and cannot represent a physical out-going on-shell photon.  The equation
of motion for $A_0$ is $\vec{\nabla}^2 A_0 =-\rho(x)$, where $\rho(x)$ is a
charge density.  If we want to allow time dependent $A_0$, then 
$\nabla^2  \partial_0 A_0 =  -\partial_0 \rho(x,t)$, 
but from current
conservation we have $\partial_0 \rho(x)=\nabla\cdot\vec{j}$ 
where $\vec{j}$ is the 3-current.
 Hence, we have  $\partial_0 A_0 = -(1/\vec{\nabla}^2 )\vec{\nabla}\cdot \vec{j} $.
 This means that if $A_0$ is to be time dependent, then there must
necessarily be a 3-current, hence there is a source
for the vector potential, $\vec{A}$, and we cannot maintain $\vec{A}=0$. 

Let us impose
the condition $\vec{\nabla}\cdot\vec{A}=0$. $\vec{A}$ 
satisfies $ (\partial_0^2-\nabla^2) \vec{A} 
-\vec{\nabla}\partial_0 A_0= \vec{j}$ ( \ie,  
$\partial_\mu F^{\mu i} = j^i$ ). 
This is often written as 
$ (\partial_0^2-\nabla^2) \vec{A} 
 = \vec{j}_T $
where $\vec{j}_T $ is the ``transverse current'' \cite{Jackson}.
Upon eliminating $\partial_0 A_0$, the transverse current
takes the nonlocal form 
$\vec{j}_T = \vec{j}-\vec{\nabla}(1/\nabla^2) \vec{\nabla}\cdot \vec{j} $.
Thus, introducing $A_0$ time dependence requires a nonzero vector potential,
and its source is essentially nonlocal. 
The nonlocal
term we obtained in eq.(\ref{three}) 
is the analogue of the transverse current \cite{Hill2}.

As stated above, the calculation in Flambaum, \etal,  was  restricted 
to a  4-vector potential of the pure Coulomb form, $A_\mu=(A_0, \vec{0})$
\ie, $\vec{E} = \vec{\nabla }A_0$. This is {\em not a gauge choice}, since
 a general 4-vector potential, $A_\mu(x,t)$, cannot be brought to the pure timelike
form by a gauge transformation, and if $\vec{A}=0$  then  $A_0$
must be static in time.  Thus a pure Coulomb potential cannot probe
an OEDM since the action averages to zero in time.

\vspace{0.1in}

In conclusion, 
Ref.\cite{Flambaum} has argued that Fig.(1) is zero. However, they have
made specific assumptions that enforce 
a static electric field configuration, and end up  computing a total
spatial divergence which is automatically null.
From this they argue that there can be
no induced effective OEDM for the electron.
However, they have not considered the case of a time dependent
radiation field, or even a homogeneous field that has a Fourier time component
matched to the oscillation frequency of the axion.
  
The diagram of Fig.(1)
represents real physics, and can be interpreted as the effective action
of an induced electron OEDM, interacting with a coherent oscillating axion field.
It produces
electric $N$-pole radiation emanating from any magnetic $N$-pole placed 
in the oscillating cosmic axion field. This can be
seen in various quantum computations at various levels
of detail \cite{Hill1,Hill2}, or directly from Maxwell's equations \cite{Hill3}.
The emission of electric
dipole radiation from magnets 
could form a basis for broadband radiative detectors for cosmic
axions. 
These conclusions have certainly not been falsified 
by the authors of ref.\cite{Flambaum}.

\vspace{0.1in}
I thank, for discussions,
Bill Bardeen, Aaron Chou,
Graham Ross, Arkady Vainshtein, and various members of the Fermilab
axion search and breakfast groups.  
This work was done at Fermilab, operated by Fermi Research Alliance, 
LLC under Contract No. DE-AC02-07CH11359 with the United States Department of Energy.

\end{document}